\documentclass{ws-procs9x6}

\begin{document}

\title{Uncovering Low-Dimensional Topological Structure 
in the QCD Vacuum}

\author{I.~HORV\'ATH$^{\lowercase{a}}$\footnote{\uppercase{S}peaker at 
the ``\uppercase{C}onfinement \uppercase{V}'' \uppercase{C}onference, 
\uppercase{G}argnano, \uppercase{I}taly, \uppercase{S}ep 10--14, 2002.},
S.J.~DONG$^{\lowercase{a}}$, T.~DRAPER$^{\lowercase{a}}$, 
K.F.~LIU$^{\lowercase{a}}$, N.~MATHUR$^{\lowercase{a}}$, 
F.X.~LEE$^{\lowercase{b,c}}$, H.B.~THACKER$^{\lowercase{d}}$  AND
J.B.~ZHANG$^{\lowercase{e}}$}

\address{
$^{a}$Department of Physics, University of Kentucky, Lexington, KY 40506 \\
$^{b}$Center for Nuclear Studies and Department of Physics,
      George Washington University, Washington, DC 20052 \\
$^{c}$Jefferson Lab, 12000 Jefferson Avenue, Newport News, VA 23606 \\
$^{d}$Department of Physics, University of Virginia, Charlottesville, VA 22901 \\
$^{e}$CSSM and Department of Physics and Mathematical Physics, 
      University of Adelaide, Adelaide, SA 5005, Australia}


\maketitle

\abstracts{
Recently, we have pointed out that sign-coherent 4-dimensional structures can 
not dominate topological charge fluctuations in QCD vacuum at all scales. 
Here we show that an enhanced lower-dimensional coherence is possible. 
In pure SU(3) lattice gauge theory we find that in a typical equilibrium 
configuration about 80\% of space-time points are covered by two 
oppositely-charged connected structures built of elementary 3-dimensional 
coherent hypercubes. The hypercubes within the structure are connected 
through 2-dimensional common faces. We suggest that this coherence is 
a manifestation of a low-dimensional order present in the QCD vacuum. 
The use of a topological charge density associated with Ginsparg-Wilson 
fermions {\it (``chiral smoothing'')} is crucial for observing this structure.
}

An important aspect of the QCD vacuum structure relates to the local patterns 
in topological charge fluctuations. Phenomena such as a large $\eta'$ mass, 
the $\theta$-dependence, and possibly also spontaneous chiral symmetry 
breaking (S$\chi$SB) are directly related to these fluctuations. Contrary to 
the situation in the past, the latest developments related to lattice chiral 
symmetry allow us to study these issues without a bias related to a particular
model picture. This approach was initiated in Ref.~\cite{Hor02B} and is based
on the use of the topological charge density 
$q_x = -\mbox{\rm tr} \,\gamma_5 \, (1 - \frac{1}{2}D_{x,x})$
associated with Ginsparg-Wilson fermions~\cite{Has98A}. This density 
(a) satisfies the index theorem on the lattice~\cite{Has98A}, 
(b) its renormalization properties are analogous to those in the 
continuum~\cite{Giu01A}, 
(c) exhibits smoother behavior than the underlying gauge fields due 
to non-ultralocality ({\it``chiral smoothing''})~\cite{Hor02B,Hor02A}, and 
(d) can be naturally eigenmode-expanded~\cite{Hor02B}. 

Due to (d) it is possible to define an {\it effective density}~\cite{Hor02B} 
encoding the topological charge fluctuations up to low-energy scale $\Lambda$, 
i.e. $q_x^{(\Lambda)} \,\equiv\, 
   -\sum_{|\lambda|\le\Lambda a} (1 - \frac{\lambda}{2})\, c^{\lambda}_x,$ 
where $c^\lambda_x = \psi^{\lambda \,+}_x \gamma_5 \psi^{\lambda}_x$ is 
the local chirality of the mode with eigenvalue $\lambda$. Using this
framework it was demonstrated unambiguously that the topological charge 
fluctuations at low energy are not dominated by coherent unit lumps (e.g.
instantons), and some interesting quantitative information about the 
typical amounts of charge in the local ``hot spots'' was 
obtained~\cite{Hor02B}.

Motivated by (c), we now concentrate on the possibility that ordered behavior 
exists in the full $q_x$ containing all fluctuations up to the lattice 
cutoff. Indeed, if one is interested in the fundamental structure in the QCD 
vacuum relevant at all scales, then the full $q_x$ has to be examined. 
On the other hand, this strategy might appear rather futile since in 
the continuum~\cite{SeSt} $\langle q(x) q(0)\rangle \le 0 \;, |x|>0\,$, and thus a 4-d 
coherence associated with objects of finite physical size should not be 
prevalent in $q_x$~\cite{Hor02B}. However, such arguments do not exclude 
the dominance of an ordered coherent structure carried by lower-dimensional 
manifolds embedded in the 4-d Euclidean space. We demonstrate that a long-range 
coherence of this type exists.

If physically relevant 4-d coherent structures existed, they could 
be identified by finding all elementary sign-coherent 4-d hypercubes 
on the lattice and determining all connected regions formed by such cubes. 
Consistency of building a 4-d manifold requires that different hypercubes 
be connected through common 3-d faces. We have done such analysis using 
the overlap operator~\cite{Neu98A} and Wilson gauge action at two lattice 
spacings ($a=0.110$ fm and $a=0.082$ fm). The isolated islands of such 
coherence were found shrinking to mere points in the continuum limit 
as expected~\cite{Hor02E}. Proceeding to the lower-dimensional manifolds, 
we have applied the same procedure using 3-d elementary cubes with structures 
defined by maximal connected regions (cubes connected through 2-d faces). 
For a given configuration, we have found all such structures and ordered them 
by the number of space-time points they contain. The typical (very stable) 
situation is shown in Fig.1 (left) where we plot the fraction 
of total number of space-time points taken by a given structure. We also show 
(right) what happens for the same configuration after performing a random 
permutation of sites.

\begin{figure*}[htb!]
\centerline{
\includegraphics[width=6.3cm]{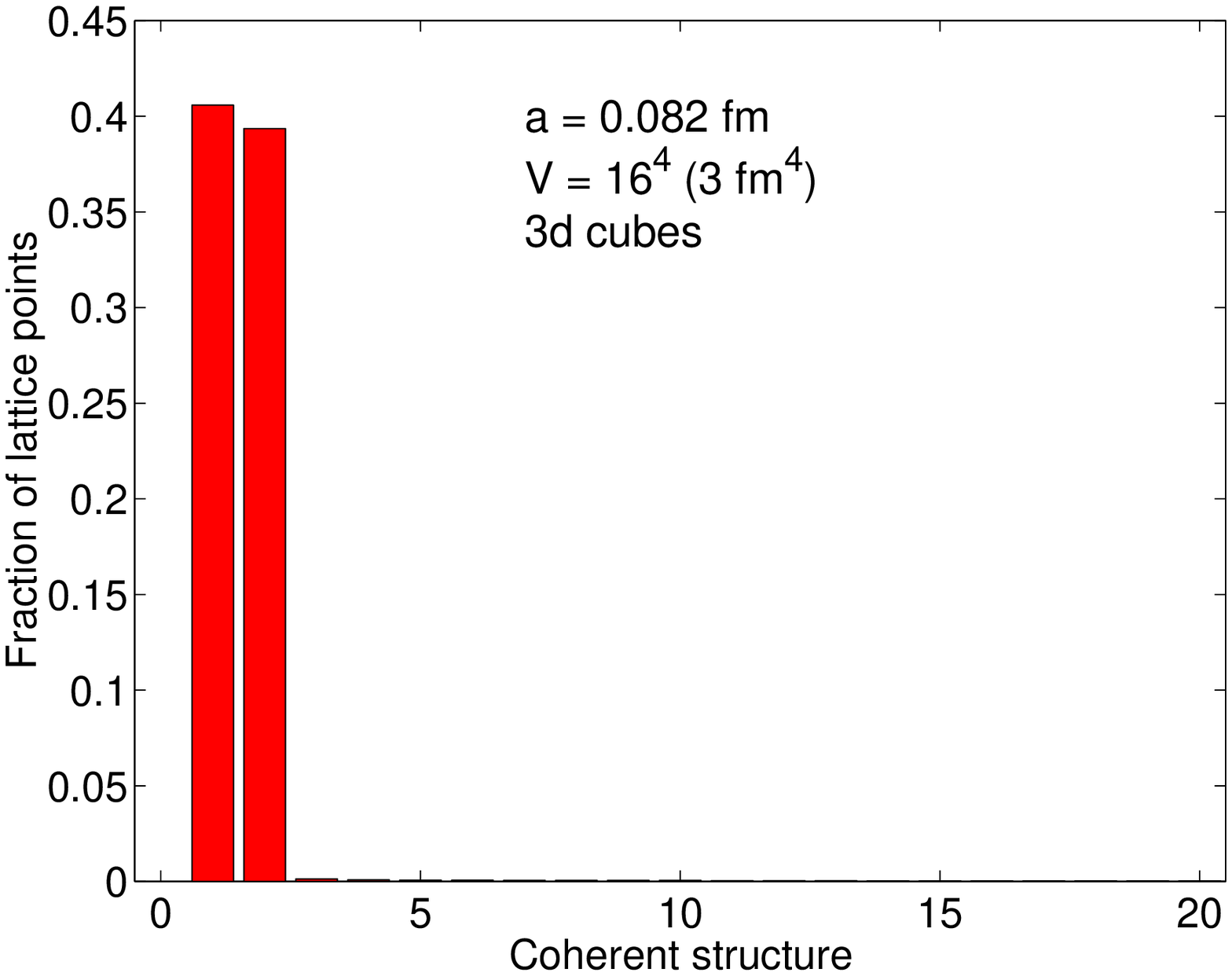}
\hspace*{-0.32in}
\includegraphics[width=6.3cm]{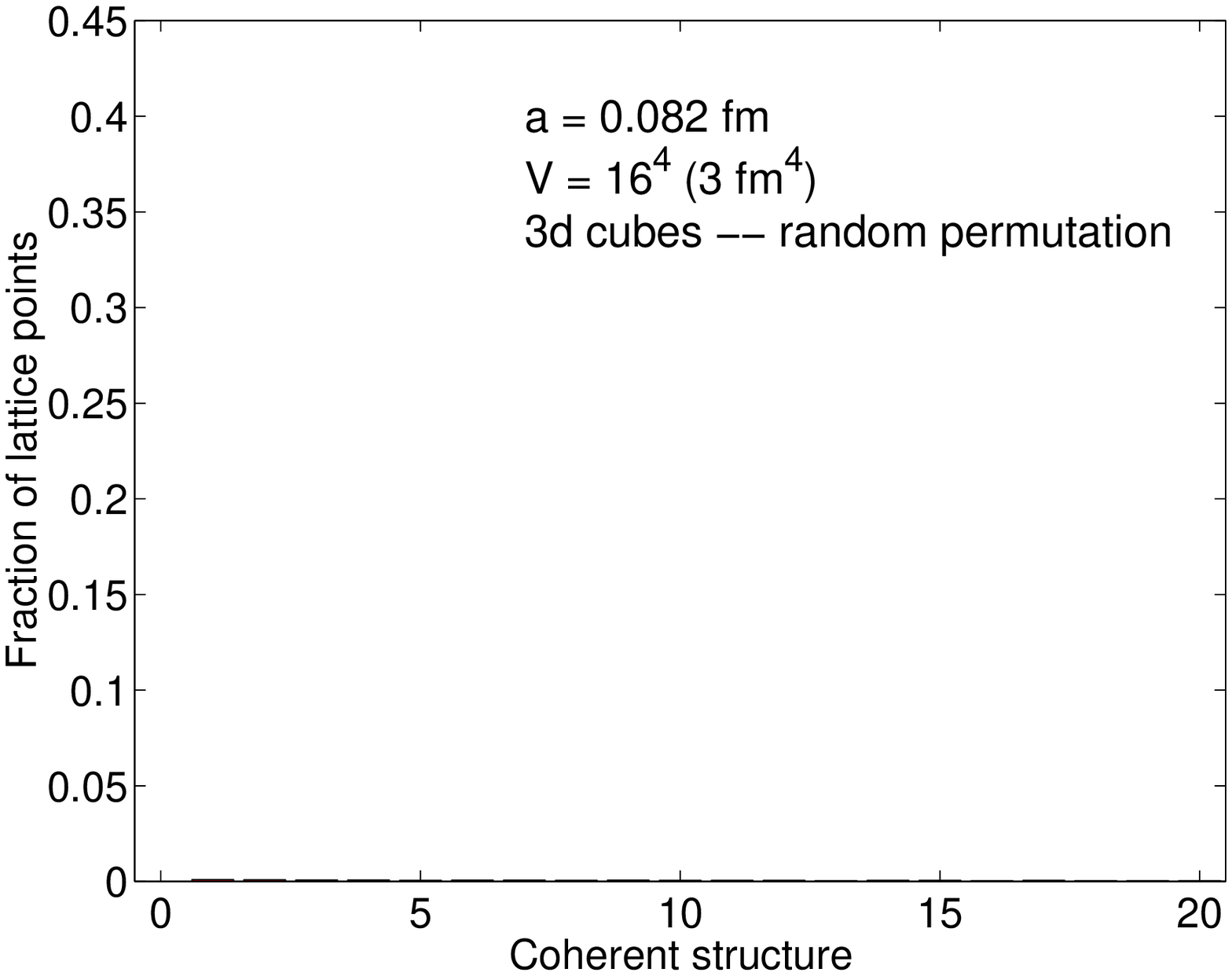}
}
\caption{Connected structures built of 3-d coherent hypercubes (see text).} 
\end{figure*}

We learn that: (i) $q_x$ is dominated by two oppositely-charged connected 
coherent structures, covering together about 80\% of space-time points 
(even after the continuum limit extrapolation). (ii) The random permutation 
of sites destroys this long-distance coherence. We thus conclude that the
observed behavior is a manifestation of an inherent low-dimensional order 
present in a typical snapshot of the QCD vacuum (equilibrium configuration) 
and absent in a completely disordered medium.

Finally, we wish to emphasize the following points. (A) From the lattice 
point of view, the two dominating structures represent 3-d hypersurfaces 
that are intertwined, embedded in the 4-d lattice, and almost filling 
the space-time. However, this is not sufficient to conclude that inside
such sheets there are 3-d coherent regions of finite physical size surviving
the continuum limit. The nature and dimensionality of such low-dimensional 
regions will be considered elsewhere~\cite{Hor02E}. (B) The use of $q_x$
associated with GW fermions {\it (chiral smoothing)} is crucial for uncovering 
this coherent structure. Indeed, simple ultralocal (even improved) gauge operators 
do not appear to exhibit such behavior. (C) We speculate that the long-range 
nature of this low-dimensional topological structure might have crucial 
consequences for understanding the mechanism of S$\chi$SB in QCD~\cite{Hor02E}.


\begin{thebibliography}{0}

\bibitem{Hor02B} I.~Horv\'ath \emph{et al}., {\tt hep-lat/0203027}.

\bibitem{Has98A}
   P. Hasenfratz, V. Laliena, F. Niedermayer,
   Phys. Lett. {\bf B427} (1998) 125.

\bibitem{Giu01A}
   L.~Giusti, G.C.~Rossi, M.~Testa, G.~Veneziano, 
   Nucl. Phys. {\bf B628} (2002) 234.

\bibitem{Hor02A} 
   I. Horv\'ath \emph{et al}., Phys. Rev. {\bf D66}, 034501 (2002).

\bibitem{SeSt}
   E. Seiler, I.O. Stamatescu, {\tt MPI-PAE/Pth 10/87}.

\bibitem{Neu98A}
   H.~Neuberger, Phys.~Lett. {\bf B427} (1998) 353.

\bibitem{Hor02E}
   I. Horv\'ath \emph{et al}., in preparation.   

\end{thebibliography}
\end{document}